\documentclass[aip,jcp,reprint]{revtex4-1}
\setcitestyle{super}

\usepackage{graphicx}% Include figure files

\usepackage{amsfonts, amsmath, amssymb,latexsym}

\usepackage{dcolumn}% Align table columns on decimal point

\usepackage{bm}% bold math

\newcommand{\me}{\mathrm{e}}

\begin{document}

\title{Quasi-Chemical Theory with Cluster Sampling from \emph{Ab Initio} 
Molecular Dynamics:  Fluoride (F$^-$) Anion Hydration}
\author{A. Muralidharan} 
\email{amuralid@tulane.edu}
\affiliation{Department of Chemical and Biomolecular Engineering, Tulane University, New Orleans, LA 70118}
\author{L. R. Pratt} 
\email{lpratt@tulane.edu}
\affiliation{Department of Chemical and Biomolecular Engineering, Tulane University, New Orleans, LA 70118}
\author{M. I. Chaudhari} 
\email{michaud@sandia.gov}
\affiliation{Sandia National Laboratories, Center for Biological and
Engineering Sciences, Albuqueruque, 87185, USA}
\author{S. B. Rempe} 
\email{slrempe@sandia.gov}
\affiliation{Sandia National Laboratories, Center for Biological and
Engineering Sciences, Albuquerque, 87185, USA}

\date{\today}

\begin{abstract} Accurate predictions of the hydration free energy for
anions  typically have been more challenging than for cations. Hydrogen bond
donation to the anion in hydrated clusters such as
$\mathrm{F(H_2O)}_n{}^-$ can lead to delicate structures. Consequently, the energy
landscape contains many local minima, even for small clusters, and these minima
present a challenge for computational optimization.  Utilization of
cluster experimental results for the free energies of gas phase clusters
shows that, even though anharmonic effects are interesting, they need not be troublesome magnitudes for careful  applications of quasi-chemical theory  to ion hydration.
Energy-optimized cluster structures for anions can leave the
central ion highly exposed and application of implicit solvation models to these structures
can incur more serious errors than for metal cations. Utilizing cluster
structures sampled from \emph{ab initio} molecular dynamics simulations
substantially fixes those issues. 
\end{abstract}

\maketitle

\section{Introduction}

Molecular quasi-chemical theory (QCT)\cite{redbook,PaulaitisPratt2002,Beck:2006wp,pratt2007potential,% 
Asthagiri:2010,Rogers:2011,Rogers} was deliberately developed from 
molecular statistical thermodynamic theory, and applications have been
both simple and remarkably accurate. This situation confronts the
canon\cite{chandler1983van} of the theory of dense liquids, and begs the
question of what accuracy may be achieved as initial simplifications are
addressed.  The present paper takes up the questions of accuracy of QCT
when refined implementations are pursued. Beyond technical theoretical
problems, this program brings forward basic questions of operational
single-ion free energies underlying rational plans to study specific ion
effects. 

The basic status of QCT may be supported, but also somewhat camouflaged,
by the fact that QCT can be closely coordinated with --- indeed
implemented through --- molecular simulation calculations. \cite{tomar2013solvation,tomar2014conditional,tomar2015importance,%
asthagiri2017intramolecular,tomar2018solvophobic} 
Earlier works termed that approach `direct' QCT.\cite{dsabo08,Jiao:co2,%
Chaudhari:Kr,Chaudhari:utility}
On that simulation basis,
QCT provides a compelling molecular theory of liquid water
itself.\cite{asthagiri2003free,paliwal2006analysis,shah2007balancing,%
chempath2008distribution,chempath2009quasichemical,weber2010molecular,doi:10.1063/1.3499315,%
doi:10.1063/1.3660205,doi:10.1063/1.3572058}  
Nevertheless, the initial
motivation\cite{redbook,martin1998hydrolysis,rempe2000hydration} 
was the
%direct 
exploitation of molecular electronic structure calculations 
in statistical thermodynamic modeling. That approach is  called `cluster' QCT. 

One motivation for the refinements (below) to cluster QCT is to achieve operational
experimental testing of the resulting free
energies.\cite{friedman1973thermodynamics}  A practical consequence of
that goal is that the QCT should be applied to both cation and anion
hydration cases.\cite{Beck:2013gp,ACR} Cations interact with ligating atoms with partially negative charge, 
like oxygens from waters. Recent studies yielded new insights on hydration,\cite{rempe2000hydration,Rempe:2001,Rempe:K,%
Asthagiri:divalents,Jiao:2011,Alam:Li,Sabo:2013gs,Mason:Li,Chaudhari:Ba}, mechanisms of selective ion binding,\cite{Varma:2007ej,Varma:2008kl,Varma:2008fh,Varma:2010,%
Varma:2011ho,Rossi:2013gm,Chaudhari:Sr} and specific ion effects for cations.\cite{Stevens}  But anion hydration clusters  
often exhibit H-bond donation to the ion.  Those clusters can be
structurally delicate, making hydrated anions  more
challenging cases.\cite{chaudhari2017quasi}  In the work below, we treat
LiF(aq) for the desired testing.  
For clarity of exposition, we
refer to F$^-$(aq) or Li$^+$(aq) when a generic single ion is discussed.

The case of F$^-$(aq) fits the description above.  An initial QCT
application works simply with reasonable accuracy.\cite{chaudhari2017quasi} Nevertheless,
refinement of that initial application requires consideration of further
technical issues that are taken up here; specifically, quantification of
anharmonic effects on free energies of hydrated ion clusters, and the
sufficiency of the polarizable continuum model (PCM) for the hydration
free energy of those clusters.

\subsection*{QCT Basics}
QCT treats the ion-water clusters as molecular species of
the system under analysis,\cite{redbook,PaulaitisPratt2002,Beck:2006wp,pratt2007potential,%
Asthagiri:2010,Rogers:2011,Rogers}
then provides  a concise format 
\begin{multline}
\mu^{\mathrm{(ex)}}_{\mathrm{F}^{-}} = -RT\ln K^{(0)}_{n}\rho_{\mathrm{H_2O}}{}^{n}
		+RT\ln p_{\mathrm{F}^{-}}(n) \\ 
	+\left(\mu^{\mathrm{(ex)}}_{\mathrm{F(H_2O)}_n{}^{-}}-n\mu^{\mathrm{(ex)}}_{\mathrm{H_2O}}\right)~,
		\label{eq:1}
\end{multline}
for free energies of solution
components such as F$^-$(aq).  
The populations of the clusters
are established by applying a clustering algorithm,
according to which proximal ligands of a specific ion are defined as 
\emph{inner-shell} partners of that ion.
The chemical association process,
\begin{equation} 
 n \mathrm{H_2O}  + \mathrm{F}^{-} \rightleftharpoons \mathrm{F(H_2O)}_n{}^{-}~,
\label{eq:2charged}
\end{equation}
introduces the equilibrium ratio,
\begin{eqnarray}
K_n =
\frac{p_{{\mathrm{F}^{-}}}(n)}{\rho_{\mathrm{H_2O}}{}^n p_{{\mathrm{F}^{-}}}(0)}~,
\label{eq:Kratio}
\end{eqnarray}
with $p_{{\mathrm{F}^{-}}}(n)$ as the thermal probability that a
specific ion has $n$ inner-shell partners.  The factor $K^{(0)}_{n}$
appearing in Eq.~\eqref{eq:1} is the equilibrium constant $K_n$
evaluated for the case that the external medium is an ideal gas. 
Evaluation of $K^{(0)}_{n}$ is accessible with widely available tools of
few-body molecular theoretical chemistry. 

Application of QCT to F$^-$(aq) thus begins with
identification of inner-shell configurations of the medium relative to
the ion.  Emphasizing the observation above that the challenge  in
treating anions lies in the variety of H-bond donation structures of
$\mathrm{F(H_2O)}_n{}^-$, a natural procedure is to
identify water molecules with H atoms within a distance of $\lambda$ from
F$^-$ as clustered; that is, 
as inner-shell partners with the
distinguished F$^-$ ion.  From there, with $n$ water ligands in the
cluster and directly interacting with the ion, the free energy is computed using Eq.~\eqref{eq:1}. 

For Li$^+$(aq), in contrast to F$^-$(aq), it is natural to identify
water molecules with O atoms within a given specific distance from Li$^+$ as clustered. 

The QCT formula (Eq.~\eqref{eq:1}) is correct for any physical choices of
$\lambda$ and $n$. We emphasize that the combination of terms on the
right-side should be independent of $n$.  This leads to two further
observations. Firstly, for intuitive choices 
of $\lambda$, the formula Eq.~\eqref{eq:1} provides a theory of the
populations $p_{{\mathrm{F}^{-}}}(n)$ to within a  constant, the
left-side of Eq.~\eqref{eq:1}; that is,  
to within \emph{post hoc}
normalization. The most probable value of $n$
provides the minimum value of $\ln p_{{\mathrm{F}^{-}}}(n)$, and it is
then natural to simply drop that contribution.  

Secondly, the independence of Eq.~\eqref{eq:1}  on $n$ provides an
indication of the sufficiency of operational approximations adopted to
evaluate the several contributions.

For the leading contribution,  $K^{(0)}_{n}$ of Eq.~\eqref{eq:1}, a
harmonic approximation for that isolated cluster contribution has
 been convenient and typically satisfactory. 
 But $K^{(0)}_{n}$ is a characteristic of
few-body molecular cluster chemistry, and is available from separate
cluster experiments.\cite{tissandier1998proton}  Those experimental
results permit the work here to avoid that harmonic approximation.  In
this way, the evaluations below treat anharmonic motion of the clusters
considered, and specifically anharmonic zero-point motion, thus further
quantifying the significance of the harmonic approximation.

In discussing the various contributions to Eq.~\eqref{eq:1}, we arrive
finally at the right-most contribution
$\left(\mu^{\mathrm{(ex)}}_{\mathrm{F(H_2O)}_n{}^{-}}-n\mu^{\mathrm{(ex)
}}_{\mathrm{H_2O}}\right)$. Here, we will adopt the polarizable dielectric
continuum model (PCM).\cite{Tomasi:2005tc} Although PCM is an extreme approximation,
the supporting physical arguments are that electrostatic effects on
chemical energy scales clearly dominate more subtle statistical
mechanical issues.  Moreover, sensitivities to artificial parameters of
the PCM, specifically the radii that establish the boundary of the
dielectric continuum, are moderated in forming the required difference. 
Nevertheless, the PCM approximation applied in this combination is the
principal approximation of the work below. That again highlights the challenge
of treating anions such as F$^-$.  Specifically, if H-bond donation leads to exotic
structures for the isolated $\mathrm{F(H_2O)}_n{}^-$ clusters with the
ion core highly exposed, that might be a problematic circumstance for the
PCM approach.  To address this possibility, the work below carries out an
inverse sampling approach exploiting the structural results from $ab$ $initio$ molecular dynamics (AIMD) simulations.  Those
details will be discussed, along with other methods, in the following section.

\begin{figure}
\includegraphics[width=3.5in]{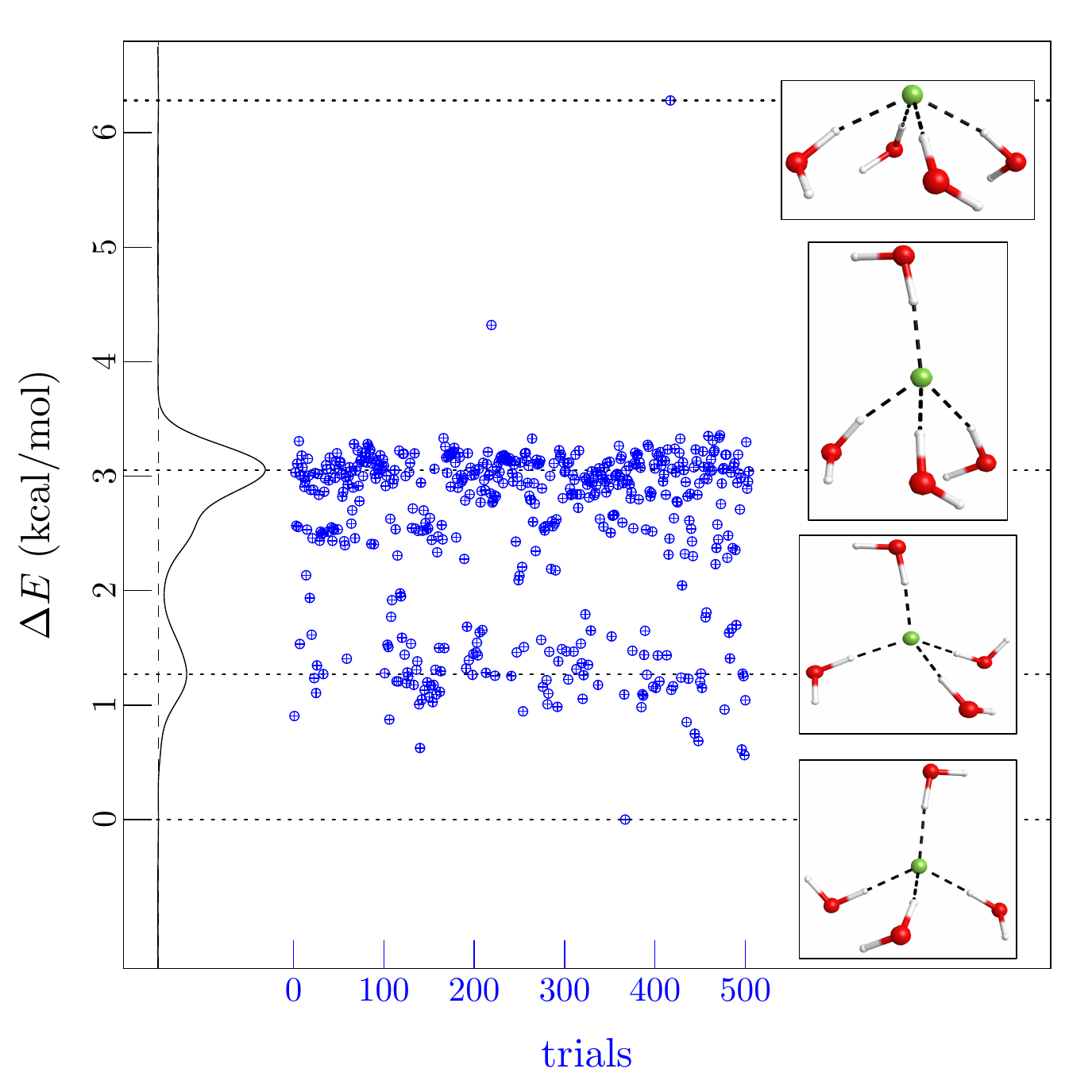}
\caption{Electronic energy of the optimized $n=4$ clusters (blue dots) starting with
configurations sampled from \emph{ab initio} molecular dynamics. The black curve shows the 
distribution of these energies.  The
lowest energy optimum (bottom inset) is about 6 kcal/mol lower in energy
than the highest energy optimum (top inset). Harmonic approximation
applied to that lowest energy cluster is in good agreement with cluster
experiments (FIG.~\ref{fig:cluster_form}).}
\label{fig:sample_N4} \end{figure}

\section{\label{sec:level1}Methods} 

\subsection*{Electronic structure calculations and cluster analysis}

All electronic structure calculations were carried out using Gaussian 09 (Rev.
D.01).\cite{frisch1998gaussian} Geometry optimizations were performed on the
$\mathrm{F(H_2O)}_n{}^-$ cluster with 1000 initial cluster configurations
sampled from AIMD. The UPBE1PBE\cite{perdew1996generalized} and
B3LYP\cite{becke1993density,lee1988development} hybrid density functionals were
utilized with the aug-cc-pVDZ\cite{dunning1989gaussian} basis set. Normal mode 
analysis was then carried out on all 1000 optimized structures to determine
zero-point corrected energies and vibrational frequencies.\cite{Rempe:normal} The absence of
imaginary frequencies confirmed that all optimizations resulted in true energy
minima.

These calculations also evaluate the single molecule (or cluster)
vibrational/rotational partition functions and thus $K^{(0)}_{n}$.  
Here, the electronic structure calculations analyze
rotations of a specific optimized structure without respect to symmetry.
Thus, the free energy integrations require some discussion of symmetry
numbers for these
molecules.\cite{Ochterski2000thermochemistry,Irikura:1998te}  The
symmetry number of 2 for the H$_2$O ligands is elementary
and included by hand in our results below.  

Considering next the molecular
cluster, the general formula for $K^{(0)}_{n}$ presents a $1/n!$
factor,\cite{redbook,Beck:2006wp} since the $n$ ligands are treated
identically.  This factor compensates for the coverage of the ligand
conformational space in a general configurational integration.  In
contrast, the electronic structure calculations exploited here integrate over rotations of a
specific cluster structure. The rotational partition functions for a
cluster structure might be expected to accomplish, as a practical matter,
interchange of the chemically identical ligands. But considering, for
example, the case $n$ = 4, rigid body rotations do not achieve
inversions of a general structure.  Thus, the full permutation group of
order $n!$ = 24 is not recovered.  Therefore, we simply supply by hand a
symmetry number of  $n!/2$, 12 rather than 24, for the case $n$ = 4. 
The case for $n$ = 5, which is considered below, is less simple than $n$=4 and goes
beyond merely prescribing the full configurational integration.  As an 
expedient, we use the value of  $n!/2$ for that case also.

\subsection*{AIMD simulations}
A system consisting of a single fluoride ion and 64 waters was simulated
using the VASP AIMD simulation package.\cite{kresse1993ab,kresse1996efficient} A cubic box of 1.2417~nm was
used to match the experimental density of water at standard conditions.
The PW91 generalized gradient approximation described the core-valence
interactions using the projector augmented-wave (PAW) method. Plane
waves with a high kinetic energy cutoff of 400~eV and a time step of
0.5~fs were used for the simulation in an NVE ensemble. A temperature of
350~K was used during the simulation run to avoid glassy behavior that can
result at lower $T$.\cite{Rempe:water}
A total simulation time of 100 ps was run and the last 50 ps trajectory was used for analysis.  

\begin{figure}
\includegraphics[width=3.5in]{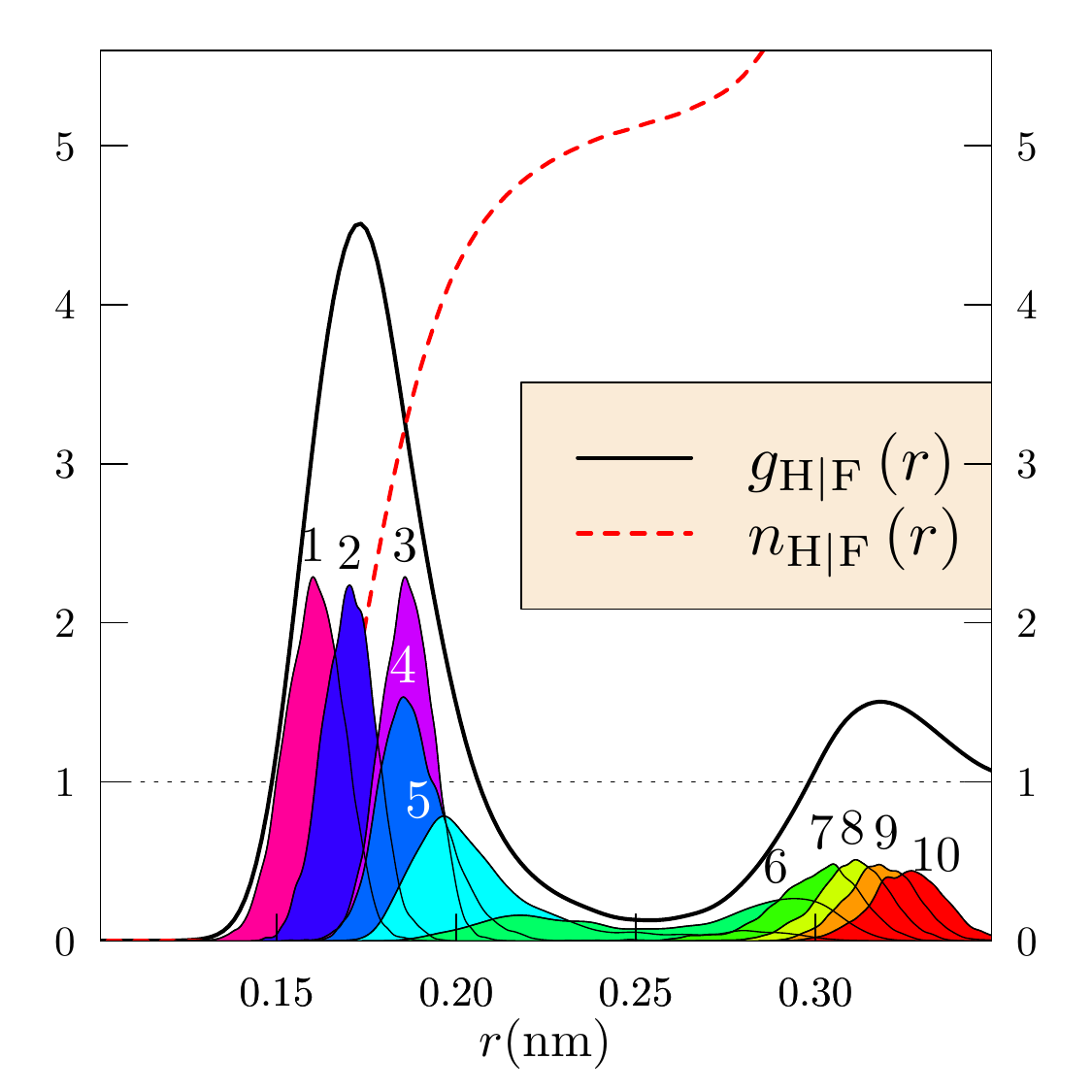}
\caption{Radial distributions of water H atoms relative to F$^-$ from
AIMD. The integer-labeled distributions are the neighborship-ordered
contributions of the $n$-th nearest H atom. A choice $\lambda \leq 0.2$
nm excludes split-shell clusters. Note the this neighborship
discrimination is sharper here when H-atoms are utilized than when
O-atoms were utilized in initial work.\cite{chaudhari2017quasi}}
\label{fig:Neighborship}
\end{figure}

\subsection*{\label{sec:level2C}PCM for AIMD sampled clusters}

The outer-shell contribution to the hydration free energy
$\left(\mu^{\mathrm{(ex)}}_{\mathrm{F(H_2O)}_n{}^{-}}-n\mu^{\mathrm{(ex)
}}_{\mathrm{H_2O}}\right)$ is treated using the polarizable continuum
model (PCM).\cite{Tomasi:2005tc}  $N_s = 1000$ $n$-cluster configurations
were extracted from the last 50~ps of AIMD simulation,  followed by single point
electronic energy calculations with PCM as the dielectric medium and separate calculations in the gas~phase.
The difference, $\varepsilon = \Delta U$, is employed in computing
\begin{eqnarray}
\mu_{\mathrm{F(H_2O)}_{n}{}^{-}}^\mathrm{(ex)} = RT
\ln \left\lbrack
\left(\frac{1}{N_s}\right)\sum\limits_{j=1}^{N_s}\me^{\varepsilon_j/RT} 
\right\rbrack .
\label{eq:rev_widom}
	\end{eqnarray}
This approach corresponds to the PDT formula\cite{Beck:2006wp,dullens2005widom}
for the inverse case, that is, %\emph{i.e.,} 
particle deletion. We explicitly
verified that these PCM results were insensitive to the specific value
of the solution dielectric constant in a high $\epsilon$ range relevant
to liquid water; the $\epsilon \rightarrow \infty$ could have been used
with imperceptible difference.  Therefore, although the configurations
sampled here correspond to AIMD trajectories at 350~K, we expect the
differences in PCM-single-point energies at 298~K to be small.
Those differences could, of course, be studied in future work. 

In PCM, the boundary around the solute is defined by spheres centered on each of
the atoms. The sensitivity of the outer-shell contribution to the size
of this cavity is characterized by changing the radius,
$R_{\mathrm{F}^{-}}$, of the solute atom between 0.15~nm and 0.20~nm.
Our operational value was
$R_{\mathrm{F}^{-}} = 0.169$~nm, the default PCM radius. That radius lies close
 to the maximum in the radial distribution 
of water hydrogens about the anion (FIG.~\ref{fig:Neighborship}). %is that the peak of g(r)?

\section{Results}
\subsection*{Clustering constraint and solution structure} 
The arrangements of ligands around a solute guide our assignment of the
clustering radius, $\lambda$. The radial distribution function of
water-H atoms relative to F$^-$,  when resolved into neighborship-ordered
contributions, clarify those considerations
(FIG.~\ref{fig:Neighborship}). Notice that the 6th-nearest neighbor
distribution is bimodal, with peaks on both sides of the minimum of
$g_{\mathrm{H\vert F}}\left(r\right)$. A choice of $\lambda\leq0.20$ nm
excludes such split-shell occupancies,\cite{chaudhari2017quasi,Sabo:2013gs} 
that is, waters that are not always in  
direct contact with the ion.
The thermal probability, $p_{\mathrm{F}^{-}}(n)$, contributing to the
second term of Eq. \eqref{eq:1}, is then evaluated based on this
constraint.

\subsection*{\label{sec:level2B}Accuracy of the harmonic approximation} 

\begin{figure}
\includegraphics[width=3.5in]{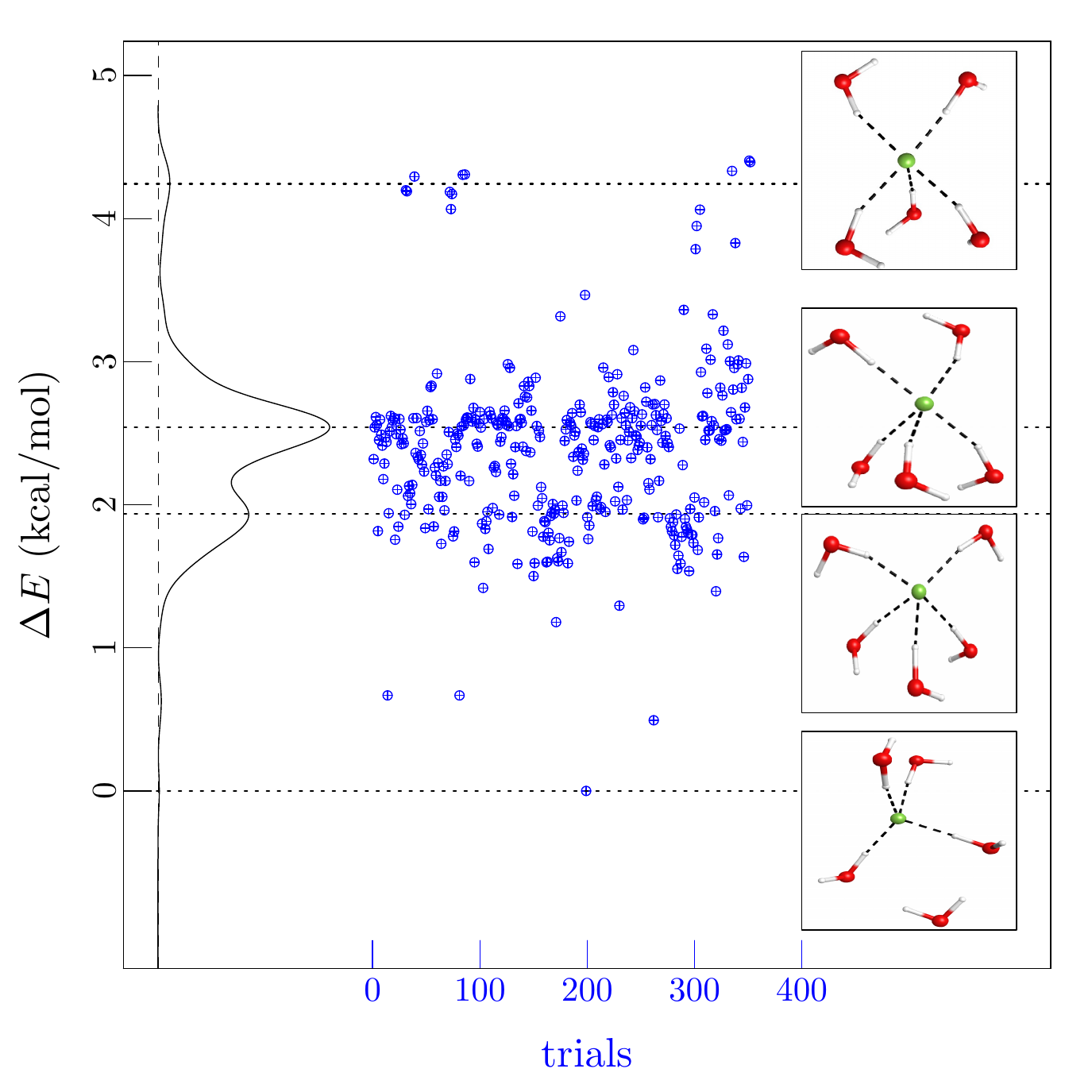}
\caption{Electronic energy of the optimized $n=5$ clusters (blue dots)
starting with configurations sampled from \emph{ab initio} molecular
dynamics. The black curve shows the distribution of these energies.  Two
kinds of optimized structures are observed here. Lowest energy,
`\emph{split-shell}' (4+1) structures (bottom inset), which do not obey
the clustering constraint, are the first kind. The lowest energy (5+0)
structures are the second kind and they conform to the clustering
constraint on which QCT theory is developed.}
\label{fig:sample_N5} \end{figure}

We extracted $N_s = 1000$ $n$-clustered configurations from the last
50~ps of AIMD simulations and optimized their geometries with respect to energy in the gas phase (FIG.
\ref{fig:sample_N4},\ref{fig:sample_N5}). Only cluster sizes with $2
\leq n \leq 5$ were observed for $\lambda=0.2$~nm. For the case of $n=4$, 
optimized energies span a range of 1-3 kcal/mol in a bimodal fashion (FIG. \ref{fig:sample_N4}).
Those energy differences are attributed to subtle
changes in the orientations of the ligands.

For $n=5$, some optimizations pushed a water molecule out to form
structures that we label henceforth as ``4+1'' (bottom inset). The rest 
are ``5+0'' optimized clusters (top inset). The ``4+1'' structures
violate the clustering constraint and are not included in evaluation
of the free energy. This issue does not arise for clusters with
$n \leq 4$.

\begin{figure}
\includegraphics[width=3.5in]{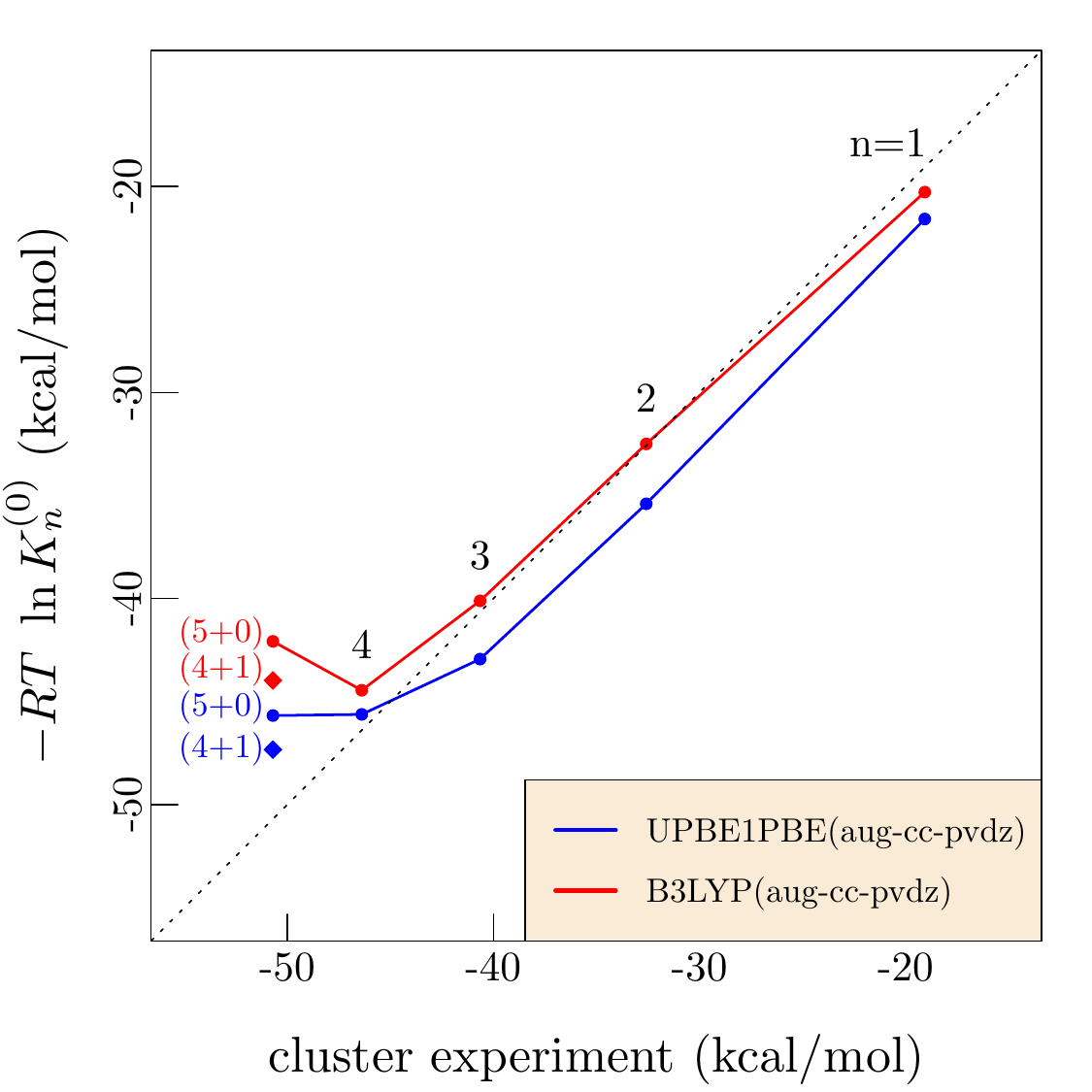}
\caption{Evaluation of $-k_{\mathrm{B}}T\ln K^{(0)}_{n}$ using the
harmonic approximation agrees well with cluster
experiments\cite{tissandier1998proton} for cluster sizes $n\leq 4$. For
$n=5$, the results suggest that (4+1) structures contribute importantly
to these experiments. Anharmonicity effects likely make less specific,
more quantitative contributions elsewhere.\cite{kathmann2007critical}
The B3LYP functional, which performs better here, is adopted for further
analyses including outer-shell contributions.} 
\label{fig:cluster_form} \end{figure}

Finally, $-RT\ln K^{(0)}_{n}$ is evaluated within a harmonic assumption for
the lowest energy structure (FIG.~\ref{fig:cluster_form}). Comparison
with cluster experiments\cite{tissandier1998proton} indicates accurate
agreement for cluster sizes $n \leq 4$. The $n=5$ case suggests that the
``4+1'' structures contribute importantly to these experiments, as also
found earlier for sodium.\cite{Soniat:2015kl} The   %cite Soniat; as also found earlier for Na
B3LYP functional, which performs better in this comparison, is adopted
for further analyses that include outer-shell contributions.

\begin{figure}
\includegraphics[width=3.5in]{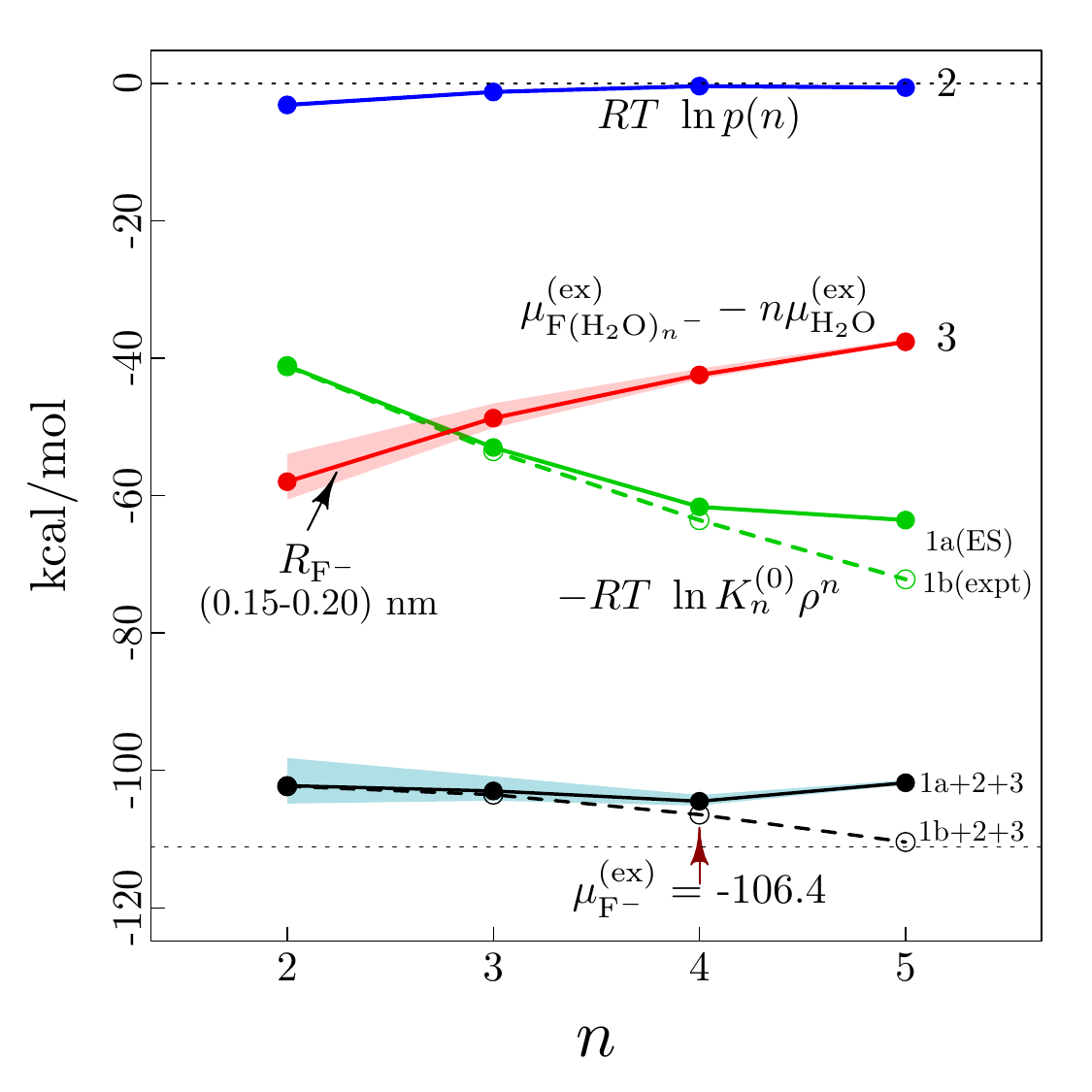}
\caption{Topmost: The free energy contribution related to cluster
poly-dispersity. Red: The outer shell contribution to the hydration free energy
evaluated using the PCM model.\cite{Tomasi:2005tc} The red band is obtained by
varying $R_{\mathrm{F}^{-}}$, as indicated in the text. Green: Isolated cluster
free energy (FIG.~\ref{fig:cluster_form}), with the corresponding dashed curve 
from cluster experiments.\cite{tissandier1998proton} Black: The
excess free energy of hydration for F$^-$,
$\mu^{\mathrm{(ex)}}_{\mathrm{F}^{-}}$, evaluated using Eq.~\eqref{eq:1}. The
sum of these contributions is substantially independent of $n$. The black dotted
line ($-111.1$ kcal/mol) is the value tabulated by
Marcus.\cite{Marcus:1994ci}} \label{fig:full_b3lyp} \end{figure}

\section{\label{sec:level2}Discussion}

Anharmonic effects on cluster free energies are easy to observe and 
interesting (FIGS.~\ref{fig:sample_N4} \& \ref{fig:sample_N5}).  Even for such a
small cluster as F(H$_2$O)$_4{}^-$, the energy landscape has many local
minima.  These minima are a challenge for 
computational optimization.  Nevertheless, comparison of 
theoretical results with experimental assessment of anharmonic 
effects, and including zero-point motion on that basis, shows that the 
differences for cluster free energies are not troublesome for this
QCT application (FIG.~\ref{fig:cluster_form}, \ref{fig:full_b3lyp}). 

The sum of the several QCT contributions for F$^-$(aq)
(FIG.~\ref{fig:full_b3lyp}) is substantially independent of $n$, which
adds confidence to the QCT results. The predicted hydration free energy
($-106.4$ kcal/mol) is in good agreement with experimental tabulation of
Marcus ($-111$~kcal/mol).\cite{Marcus:1994ci}  

A similar QCT analysis for Li$^+$(aq) arrives at a value of
$-121.1$~kcal/mol for $\mu^{\mathrm{(ex)}}_{\mathrm{Li}^{+}}$
(see Supplementary Information). Then the QCT prediction for the neutral
combination, $\mu_{\mathrm{F^{-}}}^{\mathrm{(ex)}} +
\mu_{\mathrm{Li^{+}}}^{\mathrm{(ex)}}$ = $-227.5$~kcal/mol, is in fair
agreement with experimental
tabulations, which range between $-229$~kcal/mol 
and $-232$~kcal/mol.\cite{friedman1973thermodynamics,hunenberger2011single,%
Marcus:1994ci,tissandier1998proton} Issues of a surface potential are
not involved in this comparison.\cite{Leung:2009dx,Rempe:2010er}

\section{\label{sec:level2}Conclusions}

The use of the PCM implicit solvation model is  simple and physically natural, but is ultimately
the most serious approximation for QCT applications to anions in water.
On the one hand, the simplest QCT application results in the central ion well-covered
with inner-shell ligands before relying on a traditional PCM
approximation. On the other hand, energy-optimized cluster structures
for \emph{anions} can leave the central ion highly exposed.
Application of PCM to those structures incurs more serious errors than
for metal cations. Cluster sampling from \emph{ab initio} molecular
dynamics substantially fixes that issue with the standard QCT
application. The QCT results obtained that way for LiF,  a neutral
combination so not involving a surface
potential,\cite{you2014comparison} shows fair agreement with
experimental free energies. We emphasize that we do not address here the
partial molar volumes of simple ions in water on this basis of QCT,
though the ground-work for that next challenge was laid many years
ago.\cite{redbook}

\section{Acknowledgment}
Sandia National Laboratories is a multimission laboratory managed and
operated by National Technology \& Engineering Solutions of Sandia, LLC,
a wholly owned subsidiary of Honeywell International Inc., for the U.S.
Department of Energy’s National Nuclear Security Administration under
contract DE-NA0003525. This work was supported by Sandia's LDRD program.
This work was performed, in part, at the Center for Integrated
Nanotechnologies (CINT), an Office of Science User Facility operated for
the U.S. DOE's Office of Science by Los Alamos National Laboratory
(Contract DE-AC52-06NA25296) and SNL. The views expressed in the article
do not necessarily represent the views of the U.S. Department of Energy
or the United States Government.

%\bibliography{bib,baBib,QCTbib,MypaperBib,Ca_Mg_rdf_expt,Sr_bib,Ca_Mg,Additional_refs}

%merlin.mbs aipnum4-1.bst 2010-07-25 4.21a (PWD, AO, DPC) hacked
%Control: key (0)
%Control: author (8) initials jnrlst
%Control: editor formatted (1) identically to author
%Control: production of article title (-1) disabled
%Control: page (0) single
%Control: year (1) truncated
%Control: production of eprint (0) enabled
\begin{thebibliography}{64}%
\makeatletter
\providecommand \@ifxundefined [1]{%
 \@ifx{#1\undefined}
}%
\providecommand \@ifnum [1]{%
 \ifnum #1\expandafter \@firstoftwo
 \else \expandafter \@secondoftwo
 \fi
}%
\providecommand \@ifx [1]{%
 \ifx #1\expandafter \@firstoftwo
 \else \expandafter \@secondoftwo
 \fi
}%
\providecommand \natexlab [1]{#1}%
\providecommand \enquote  [1]{``#1''}%
\providecommand \bibnamefont  [1]{#1}%
\providecommand \bibfnamefont [1]{#1}%
\providecommand \citenamefont [1]{#1}%
\providecommand \href@noop [0]{\@secondoftwo}%
\providecommand \href [0]{\begingroup \@sanitize@url \@href}%
\providecommand \@href[1]{\@@startlink{#1}\@@href}%
\providecommand \@@href[1]{\endgroup#1\@@endlink}%
\providecommand \@sanitize@url [0]{\catcode `\\12\catcode `\$12\catcode
  `\&12\catcode `\#12\catcode `\^12\catcode `\_12\catcode `\%12\relax}%
\providecommand \@@startlink[1]{}%
\providecommand \@@endlink[0]{}%
\providecommand \url  [0]{\begingroup\@sanitize@url \@url }%
\providecommand \@url [1]{\endgroup\@href {#1}{\urlprefix }}%
\providecommand \urlprefix  [0]{URL }%
\providecommand \Eprint [0]{\href }%
\providecommand \doibase [0]{http://dx.doi.org/}%
\providecommand \selectlanguage [0]{\@gobble}%
\providecommand \bibinfo  [0]{\@secondoftwo}%
\providecommand \bibfield  [0]{\@secondoftwo}%
\providecommand \translation [1]{[#1]}%
\providecommand \BibitemOpen [0]{}%
\providecommand \bibitemStop [0]{}%
\providecommand \bibitemNoStop [0]{.\EOS\space}%
\providecommand \EOS [0]{\spacefactor3000\relax}%
\providecommand \BibitemShut  [1]{\csname bibitem#1\endcsname}%
\let\auto@bib@innerbib\@empty
%</preamble>
\bibitem [{\citenamefont {Pratt}\ and\ \citenamefont {Rempe}(1999)}]{redbook}%
  \BibitemOpen
  \bibfield  {author} {\bibinfo {author} {\bibfnamefont {L.~R.}\ \bibnamefont
  {Pratt}}\ and\ \bibinfo {author} {\bibfnamefont {S.~B.}\ \bibnamefont
  {Rempe}},\ }in\ \href@noop {} {\emph {\bibinfo {booktitle} {Simulation and
  Theory of Electrostatic Interactions in Solution}}},\ \bibinfo {editor}
  {edited by\ \bibinfo {editor} {\bibfnamefont {G.}~\bibnamefont {Hummer}}\
  and\ \bibinfo {editor} {\bibfnamefont {L.~R.}\ \bibnamefont {Pratt}}}\
  (\bibinfo  {publisher} {AIP: New York},\ \bibinfo {year} {1999})\ pp.\
  \bibinfo {pages} {177--201}\BibitemShut {NoStop}%
\bibitem [{\citenamefont {Paulaitis}\ and\ \citenamefont
  {Pratt}(2002)}]{PaulaitisPratt2002}%
  \BibitemOpen
  \bibfield  {author} {\bibinfo {author} {\bibfnamefont {M.~E.}\ \bibnamefont
  {Paulaitis}}\ and\ \bibinfo {author} {\bibfnamefont {L.~R.}\ \bibnamefont
  {Pratt}},\ }\href@noop {} {\bibfield  {journal} {\bibinfo  {journal} {Adv.
  Prot. Chem.}\ }\textbf {\bibinfo {volume} {62}},\ \bibinfo {pages} {283}
  (\bibinfo {year} {2002})}\BibitemShut {NoStop}%
\bibitem [{\citenamefont {Beck}, \citenamefont {Paulaitis},\ and\ \citenamefont
  {Pratt}(2006)}]{Beck:2006wp}%
  \BibitemOpen
  \bibfield  {author} {\bibinfo {author} {\bibfnamefont {T.~L.}\ \bibnamefont
  {Beck}}, \bibinfo {author} {\bibfnamefont {M.~E.}\ \bibnamefont {Paulaitis}},
  \ and\ \bibinfo {author} {\bibfnamefont {L.~R.}\ \bibnamefont {Pratt}},\
  }\href@noop {} {\emph {\bibinfo {title} {The Potential Distribution Theorem
  and Models of Molecular Solutions}}}\ (\bibinfo  {publisher} {Cambridge
  University Press},\ \bibinfo {year} {2006})\BibitemShut {NoStop}%
\bibitem [{\citenamefont {Pratt}\ and\ \citenamefont
  {Asthagiri}(2007)}]{pratt2007potential}%
  \BibitemOpen
  \bibfield  {author} {\bibinfo {author} {\bibfnamefont {L.~R.}\ \bibnamefont
  {Pratt}}\ and\ \bibinfo {author} {\bibfnamefont {D.}~\bibnamefont
  {Asthagiri}},\ }in\ \href@noop {} {\emph {\bibinfo {booktitle} {Free Energy
  Calculations}}}\ (\bibinfo  {publisher} {Springer},\ \bibinfo {year} {2007})\
  pp.\ \bibinfo {pages} {323--351}\BibitemShut {NoStop}%
\bibitem [{\citenamefont {Asthagiri}\ \emph {et~al.}(2010)\citenamefont
  {Asthagiri}, \citenamefont {Dixit}, \citenamefont {Merchant}, \citenamefont
  {Paulaitis}, \citenamefont {Pratt}, \citenamefont {Rempe},\ and\
  \citenamefont {Varma}}]{Asthagiri:2010}%
  \BibitemOpen
  \bibfield  {author} {\bibinfo {author} {\bibfnamefont {D.}~\bibnamefont
  {Asthagiri}}, \bibinfo {author} {\bibfnamefont {P.~D.}\ \bibnamefont
  {Dixit}}, \bibinfo {author} {\bibfnamefont {S.}~\bibnamefont {Merchant}},
  \bibinfo {author} {\bibfnamefont {M.~E.}\ \bibnamefont {Paulaitis}}, \bibinfo
  {author} {\bibfnamefont {L.~R.}\ \bibnamefont {Pratt}}, \bibinfo {author}
  {\bibfnamefont {S.~B.}\ \bibnamefont {Rempe}}, \ and\ \bibinfo {author}
  {\bibfnamefont {S.}~\bibnamefont {Varma}},\ }\href@noop {} {\bibfield
  {journal} {\bibinfo  {journal} {Chem. Phys. Lett.}\ }\textbf {\bibinfo
  {volume} {485}},\ \bibinfo {pages} {1} (\bibinfo {year} {2010})}\BibitemShut
  {NoStop}%
\bibitem [{\citenamefont {Rogers}\ and\ \citenamefont
  {Rempe}(2011)}]{Rogers:2011}%
  \BibitemOpen
  \bibfield  {author} {\bibinfo {author} {\bibfnamefont {D.~M.}\ \bibnamefont
  {Rogers}}\ and\ \bibinfo {author} {\bibfnamefont {S.~B.}\ \bibnamefont
  {Rempe}},\ }\href@noop {} {\bibfield  {journal} {\bibinfo  {journal} {J.
  Phys. Chem. B}\ }\textbf {\bibinfo {volume} {115}},\ \bibinfo {pages} {9116}
  (\bibinfo {year} {2011})}\BibitemShut {NoStop}%
\bibitem [{\citenamefont {Rogers}\ \emph {et~al.}(2013)\citenamefont {Rogers},
  \citenamefont {Jiao}, \citenamefont {Pratt},\ and\ \citenamefont
  {Rempe}}]{Rogers}%
  \BibitemOpen
  \bibfield  {author} {\bibinfo {author} {\bibfnamefont {D.~M.}\ \bibnamefont
  {Rogers}}, \bibinfo {author} {\bibfnamefont {D.}~\bibnamefont {Jiao}},
  \bibinfo {author} {\bibfnamefont {L.~R.}\ \bibnamefont {Pratt}}, \ and\
  \bibinfo {author} {\bibfnamefont {S.~B.}\ \bibnamefont {Rempe}},\ }\href@noop
  {} {\bibfield  {journal} {\bibinfo  {journal} {Ann. Rep. Comp. Chem.}\
  }\textbf {\bibinfo {volume} {8}},\ \bibinfo {pages} {71} (\bibinfo {year}
  {2013})}\BibitemShut {NoStop}%
\bibitem [{\citenamefont {Chandler}, \citenamefont {Weeks},\ and\ \citenamefont
  {Andersen}(1983)}]{chandler1983van}%
  \BibitemOpen
  \bibfield  {author} {\bibinfo {author} {\bibfnamefont {D.}~\bibnamefont
  {Chandler}}, \bibinfo {author} {\bibfnamefont {J.~D.}\ \bibnamefont {Weeks}},
  \ and\ \bibinfo {author} {\bibfnamefont {H.~C.}\ \bibnamefont {Andersen}},\
  }\href@noop {} {\bibfield  {journal} {\bibinfo  {journal} {Science}\ }\textbf
  {\bibinfo {volume} {220}},\ \bibinfo {pages} {787} (\bibinfo {year}
  {1983})}\BibitemShut {NoStop}%
\bibitem [{\citenamefont {Tomar}, \citenamefont {Asthagiri},\ and\
  \citenamefont {Weber}(2013)}]{tomar2013solvation}%
  \BibitemOpen
  \bibfield  {author} {\bibinfo {author} {\bibfnamefont {D.~S.}\ \bibnamefont
  {Tomar}}, \bibinfo {author} {\bibfnamefont {D.}~\bibnamefont {Asthagiri}}, \
  and\ \bibinfo {author} {\bibfnamefont {V.}~\bibnamefont {Weber}},\
  }\href@noop {} {\bibfield  {journal} {\bibinfo  {journal} {Biophys. J.}\
  }\textbf {\bibinfo {volume} {105}},\ \bibinfo {pages} {1482} (\bibinfo {year}
  {2013})}\BibitemShut {NoStop}%
\bibitem [{\citenamefont {Tomar}\ \emph {et~al.}(2014)\citenamefont {Tomar},
  \citenamefont {Weber}, \citenamefont {Pettitt},\ and\ \citenamefont
  {Asthagiri}}]{tomar2014conditional}%
  \BibitemOpen
  \bibfield  {author} {\bibinfo {author} {\bibfnamefont {D.~S.}\ \bibnamefont
  {Tomar}}, \bibinfo {author} {\bibfnamefont {V.}~\bibnamefont {Weber}},
  \bibinfo {author} {\bibfnamefont {B.~M.}\ \bibnamefont {Pettitt}}, \ and\
  \bibinfo {author} {\bibfnamefont {D.}~\bibnamefont {Asthagiri}},\ }\href@noop
  {} {\bibfield  {journal} {\bibinfo  {journal} {J. Phys. Chem. B}\ }\textbf
  {\bibinfo {volume} {118}},\ \bibinfo {pages} {4080} (\bibinfo {year}
  {2014})}\BibitemShut {NoStop}%
\bibitem [{\citenamefont {Tomar}\ \emph {et~al.}(2015)\citenamefont {Tomar},
  \citenamefont {Weber}, \citenamefont {Pettitt},\ and\ \citenamefont
  {Asthagiri}}]{tomar2015importance}%
  \BibitemOpen
  \bibfield  {author} {\bibinfo {author} {\bibfnamefont {D.~S.}\ \bibnamefont
  {Tomar}}, \bibinfo {author} {\bibfnamefont {V.}~\bibnamefont {Weber}},
  \bibinfo {author} {\bibfnamefont {B.~M.}\ \bibnamefont {Pettitt}}, \ and\
  \bibinfo {author} {\bibfnamefont {D.}~\bibnamefont {Asthagiri}},\ }\href@noop
  {} {\bibfield  {journal} {\bibinfo  {journal} {J. Phys. Chem. B}\ }\textbf
  {\bibinfo {volume} {120}},\ \bibinfo {pages} {69} (\bibinfo {year}
  {2015})}\BibitemShut {NoStop}%
\bibitem [{\citenamefont {Asthagiri}\ \emph {et~al.}(2017)\citenamefont
  {Asthagiri}, \citenamefont {Karandur}, \citenamefont {Tomar},\ and\
  \citenamefont {Pettitt}}]{asthagiri2017intramolecular}%
  \BibitemOpen
  \bibfield  {author} {\bibinfo {author} {\bibfnamefont {D.}~\bibnamefont
  {Asthagiri}}, \bibinfo {author} {\bibfnamefont {D.}~\bibnamefont {Karandur}},
  \bibinfo {author} {\bibfnamefont {D.~S.}\ \bibnamefont {Tomar}}, \ and\
  \bibinfo {author} {\bibfnamefont {B.~M.}\ \bibnamefont {Pettitt}},\
  }\href@noop {} {\bibfield  {journal} {\bibinfo  {journal} {J. Phys. Chem. B}\
  }\textbf {\bibinfo {volume} {121}},\ \bibinfo {pages} {8078} (\bibinfo {year}
  {2017})}\BibitemShut {NoStop}%
\bibitem [{\citenamefont {Tomar}, \citenamefont {Ramesh},\ and\ \citenamefont
  {Asthagiri}(2018)}]{tomar2018solvophobic}%
  \BibitemOpen
  \bibfield  {author} {\bibinfo {author} {\bibfnamefont {D.~S.}\ \bibnamefont
  {Tomar}}, \bibinfo {author} {\bibfnamefont {N.}~\bibnamefont {Ramesh}}, \
  and\ \bibinfo {author} {\bibfnamefont {D.}~\bibnamefont {Asthagiri}},\
  }\href@noop {} {\bibfield  {journal} {\bibinfo  {journal} {J. Chem. Phys.}\
  }\textbf {\bibinfo {volume} {148}},\ \bibinfo {pages} {222822} (\bibinfo
  {year} {2018})}\BibitemShut {NoStop}%
\bibitem [{\citenamefont {Sabo}\ \emph {et~al.}(2008)\citenamefont {Sabo},
  \citenamefont {Varma}, \citenamefont {Martin},\ and\ \citenamefont
  {Rempe}}]{dsabo08}%
  \BibitemOpen
  \bibfield  {author} {\bibinfo {author} {\bibfnamefont {D.}~\bibnamefont
  {Sabo}}, \bibinfo {author} {\bibfnamefont {S.}~\bibnamefont {Varma}},
  \bibinfo {author} {\bibfnamefont {M.~G.}\ \bibnamefont {Martin}}, \ and\
  \bibinfo {author} {\bibfnamefont {S.~B.}\ \bibnamefont {Rempe}},\ }\href@noop
  {} {\bibfield  {journal} {\bibinfo  {journal} {J. Phys. Chem. B}\ }\textbf
  {\bibinfo {volume} {112}},\ \bibinfo {pages} {867} (\bibinfo {year}
  {2008})}\BibitemShut {NoStop}%
\bibitem [{\citenamefont {Jiao}\ and\ \citenamefont {Rempe}(2011)}]{Jiao:co2}%
  \BibitemOpen
  \bibfield  {author} {\bibinfo {author} {\bibfnamefont {D.}~\bibnamefont
  {Jiao}}\ and\ \bibinfo {author} {\bibfnamefont {S.~B.}\ \bibnamefont
  {Rempe}},\ }\href@noop {} {\bibfield  {journal} {\bibinfo  {journal} {J.
  Chem. Phys.}\ }\textbf {\bibinfo {volume} {134}},\ \bibinfo {pages} {224506}
  (\bibinfo {year} {2011})}\BibitemShut {NoStop}%
\bibitem [{\citenamefont {Chaudhari}\ \emph {et~al.}(2015)\citenamefont
  {Chaudhari}, \citenamefont {Sabo}, \citenamefont {Pratt},\ and\ \citenamefont
  {Rempe}}]{Chaudhari:Kr}%
  \BibitemOpen
  \bibfield  {author} {\bibinfo {author} {\bibfnamefont {M.~I.}\ \bibnamefont
  {Chaudhari}}, \bibinfo {author} {\bibfnamefont {D.}~\bibnamefont {Sabo}},
  \bibinfo {author} {\bibfnamefont {L.~R.}\ \bibnamefont {Pratt}}, \ and\
  \bibinfo {author} {\bibfnamefont {S.~B.}\ \bibnamefont {Rempe}},\ }\href@noop
  {} {\bibfield  {journal} {\bibinfo  {journal} {J. Phys. Chem. B}\ }\textbf
  {\bibinfo {volume} {119}},\ \bibinfo {pages} {9098} (\bibinfo {year}
  {2015})}\BibitemShut {NoStop}%
\bibitem [{\citenamefont {Chaudhari}, \citenamefont {Pratt},\ and\
  \citenamefont {Rempe}(2018)}]{Chaudhari:utility}%
  \BibitemOpen
  \bibfield  {author} {\bibinfo {author} {\bibfnamefont {M.~I.}\ \bibnamefont
  {Chaudhari}}, \bibinfo {author} {\bibfnamefont {L.~R.}\ \bibnamefont
  {Pratt}}, \ and\ \bibinfo {author} {\bibfnamefont {S.~B.}\ \bibnamefont
  {Rempe}},\ }\href@noop {} {\bibfield  {journal} {\bibinfo  {journal} {Mol.
  Simul.}\ }\textbf {\bibinfo {volume} {44}},\ \bibinfo {pages} {110} (\bibinfo
  {year} {2018})}\BibitemShut {NoStop}%
\bibitem [{\citenamefont {Asthagiri}, \citenamefont {Pratt},\ and\
  \citenamefont {Kress}(2003)}]{asthagiri2003free}%
  \BibitemOpen
  \bibfield  {author} {\bibinfo {author} {\bibfnamefont {D.}~\bibnamefont
  {Asthagiri}}, \bibinfo {author} {\bibfnamefont {L.~R.}\ \bibnamefont
  {Pratt}}, \ and\ \bibinfo {author} {\bibfnamefont {J.}~\bibnamefont
  {Kress}},\ }\href@noop {} {\bibfield  {journal} {\bibinfo  {journal} {Phys.
  Rev. E}\ }\textbf {\bibinfo {volume} {68}},\ \bibinfo {pages} {041505}
  (\bibinfo {year} {2003})}\BibitemShut {NoStop}%
\bibitem [{\citenamefont {Paliwal}\ \emph {et~al.}(2006)\citenamefont
  {Paliwal}, \citenamefont {Asthagiri}, \citenamefont {Pratt}, \citenamefont
  {Ashbaugh},\ and\ \citenamefont {Paulaitis}}]{paliwal2006analysis}%
  \BibitemOpen
  \bibfield  {author} {\bibinfo {author} {\bibfnamefont {A.}~\bibnamefont
  {Paliwal}}, \bibinfo {author} {\bibfnamefont {D.}~\bibnamefont {Asthagiri}},
  \bibinfo {author} {\bibfnamefont {L.}~\bibnamefont {Pratt}}, \bibinfo
  {author} {\bibfnamefont {H.}~\bibnamefont {Ashbaugh}}, \ and\ \bibinfo
  {author} {\bibfnamefont {M.}~\bibnamefont {Paulaitis}},\ }\href@noop {}
  {\bibfield  {journal} {\bibinfo  {journal} {J. Chem. Phys.}\ }\textbf
  {\bibinfo {volume} {124}},\ \bibinfo {pages} {224502} (\bibinfo {year}
  {2006})}\BibitemShut {NoStop}%
\bibitem [{\citenamefont {Shah}\ \emph {et~al.}(2007)\citenamefont {Shah},
  \citenamefont {Asthagiri}, \citenamefont {Pratt},\ and\ \citenamefont
  {Paulaitis}}]{shah2007balancing}%
  \BibitemOpen
  \bibfield  {author} {\bibinfo {author} {\bibfnamefont {J.}~\bibnamefont
  {Shah}}, \bibinfo {author} {\bibfnamefont {D.}~\bibnamefont {Asthagiri}},
  \bibinfo {author} {\bibfnamefont {L.}~\bibnamefont {Pratt}}, \ and\ \bibinfo
  {author} {\bibfnamefont {M.}~\bibnamefont {Paulaitis}},\ }\href@noop {}
  {\bibfield  {journal} {\bibinfo  {journal} {J. Chem. Phys.}\ }\textbf
  {\bibinfo {volume} {127}},\ \bibinfo {pages} {144508} (\bibinfo {year}
  {2007})}\BibitemShut {NoStop}%
\bibitem [{\citenamefont {Chempath}\ and\ \citenamefont
  {Pratt}(2008)}]{chempath2008distribution}%
  \BibitemOpen
  \bibfield  {author} {\bibinfo {author} {\bibfnamefont {S.}~\bibnamefont
  {Chempath}}\ and\ \bibinfo {author} {\bibfnamefont {L.~R.}\ \bibnamefont
  {Pratt}},\ }\href@noop {} {\bibfield  {journal} {\bibinfo  {journal} {J.
  Phys. Chem. B}\ }\textbf {\bibinfo {volume} {113}},\ \bibinfo {pages} {4147}
  (\bibinfo {year} {2008})}\BibitemShut {NoStop}%
\bibitem [{\citenamefont {Chempath}, \citenamefont {Pratt},\ and\ \citenamefont
  {Paulaitis}(2009)}]{chempath2009quasichemical}%
  \BibitemOpen
  \bibfield  {author} {\bibinfo {author} {\bibfnamefont {S.}~\bibnamefont
  {Chempath}}, \bibinfo {author} {\bibfnamefont {L.~R.}\ \bibnamefont {Pratt}},
  \ and\ \bibinfo {author} {\bibfnamefont {M.~E.}\ \bibnamefont {Paulaitis}},\
  }\href@noop {} {\bibfield  {journal} {\bibinfo  {journal} {J. Chem. Phys.}\
  }\textbf {\bibinfo {volume} {130}},\ \bibinfo {pages} {054113} (\bibinfo
  {year} {2009})}\BibitemShut {NoStop}%
\bibitem [{\citenamefont {Weber}\ \emph {et~al.}(2010)\citenamefont {Weber},
  \citenamefont {Merchant}, \citenamefont {Dixit},\ and\ \citenamefont
  {Asthagiri}}]{weber2010molecular}%
  \BibitemOpen
  \bibfield  {author} {\bibinfo {author} {\bibfnamefont {V.}~\bibnamefont
  {Weber}}, \bibinfo {author} {\bibfnamefont {S.}~\bibnamefont {Merchant}},
  \bibinfo {author} {\bibfnamefont {P.~D.}\ \bibnamefont {Dixit}}, \ and\
  \bibinfo {author} {\bibfnamefont {D.}~\bibnamefont {Asthagiri}},\ }\href@noop
  {} {\bibfield  {journal} {\bibinfo  {journal} {J. Chem. Phys.}\ }\textbf
  {\bibinfo {volume} {132}},\ \bibinfo {pages} {204509} (\bibinfo {year}
  {2010})}\BibitemShut {NoStop}%
\bibitem [{\citenamefont {Weber}\ and\ \citenamefont
  {Asthagiri}(2010)}]{doi:10.1063/1.3499315}%
  \BibitemOpen
  \bibfield  {author} {\bibinfo {author} {\bibfnamefont {V.}~\bibnamefont
  {Weber}}\ and\ \bibinfo {author} {\bibfnamefont {D.}~\bibnamefont
  {Asthagiri}},\ }\href {\doibase 10.1063/1.3499315} {\bibfield  {journal}
  {\bibinfo  {journal} {J. Chem. Phys.}\ }\textbf {\bibinfo {volume} {133}},\
  \bibinfo {pages} {141101} (\bibinfo {year} {2010})}\BibitemShut {NoStop}%
\bibitem [{\citenamefont {Weber}, \citenamefont {Merchant},\ and\ \citenamefont
  {Asthagiri}(2011)}]{doi:10.1063/1.3660205}%
  \BibitemOpen
  \bibfield  {author} {\bibinfo {author} {\bibfnamefont {V.}~\bibnamefont
  {Weber}}, \bibinfo {author} {\bibfnamefont {S.}~\bibnamefont {Merchant}}, \
  and\ \bibinfo {author} {\bibfnamefont {D.}~\bibnamefont {Asthagiri}},\ }\href
  {\doibase 10.1063/1.3660205} {\bibfield  {journal} {\bibinfo  {journal} {J.
  Chem. Phys.}\ }\textbf {\bibinfo {volume} {135}},\ \bibinfo {pages} {181101}
  (\bibinfo {year} {2011})}\BibitemShut {NoStop}%
\bibitem [{\citenamefont {Merchant}, \citenamefont {Shah},\ and\ \citenamefont
  {Asthagiri}(2011)}]{doi:10.1063/1.3572058}%
  \BibitemOpen
  \bibfield  {author} {\bibinfo {author} {\bibfnamefont {S.}~\bibnamefont
  {Merchant}}, \bibinfo {author} {\bibfnamefont {J.~K.}\ \bibnamefont {Shah}},
  \ and\ \bibinfo {author} {\bibfnamefont {D.}~\bibnamefont {Asthagiri}},\
  }\href {\doibase 10.1063/1.3572058} {\bibfield  {journal} {\bibinfo
  {journal} {J. Chem. Phys.}\ }\textbf {\bibinfo {volume} {134}},\ \bibinfo
  {pages} {124514} (\bibinfo {year} {2011})}\BibitemShut {NoStop}%
\bibitem [{\citenamefont {Martin}, \citenamefont {Hay},\ and\ \citenamefont
  {Pratt}(1998)}]{martin1998hydrolysis}%
  \BibitemOpen
  \bibfield  {author} {\bibinfo {author} {\bibfnamefont {R.~L.}\ \bibnamefont
  {Martin}}, \bibinfo {author} {\bibfnamefont {P.~J.}\ \bibnamefont {Hay}}, \
  and\ \bibinfo {author} {\bibfnamefont {L.~R.}\ \bibnamefont {Pratt}},\
  }\href@noop {} {\bibfield  {journal} {\bibinfo  {journal} {J. Phys. Chem. A}\
  }\textbf {\bibinfo {volume} {102}},\ \bibinfo {pages} {3565} (\bibinfo {year}
  {1998})}\BibitemShut {NoStop}%
\bibitem [{\citenamefont {Rempe}\ \emph {et~al.}(2000)\citenamefont {Rempe},
  \citenamefont {Pratt}, \citenamefont {Hummer}, \citenamefont {Kress},
  \citenamefont {Martin},\ and\ \citenamefont {Redondo}}]{rempe2000hydration}%
  \BibitemOpen
  \bibfield  {author} {\bibinfo {author} {\bibfnamefont {S.~B.}\ \bibnamefont
  {Rempe}}, \bibinfo {author} {\bibfnamefont {L.~R.}\ \bibnamefont {Pratt}},
  \bibinfo {author} {\bibfnamefont {G.}~\bibnamefont {Hummer}}, \bibinfo
  {author} {\bibfnamefont {J.~D.}\ \bibnamefont {Kress}}, \bibinfo {author}
  {\bibfnamefont {R.~L.}\ \bibnamefont {Martin}}, \ and\ \bibinfo {author}
  {\bibfnamefont {A.}~\bibnamefont {Redondo}},\ }\href@noop {} {\bibfield
  {journal} {\bibinfo  {journal} {J. Am. Chem. Soc.}\ }\textbf {\bibinfo
  {volume} {122}},\ \bibinfo {pages} {966} (\bibinfo {year}
  {2000})}\BibitemShut {NoStop}%
\bibitem [{\citenamefont {Friedman}\ and\ \citenamefont
  {Krishnan}(1973)}]{friedman1973thermodynamics}%
  \BibitemOpen
  \bibfield  {author} {\bibinfo {author} {\bibfnamefont {H.}~\bibnamefont
  {Friedman}}\ and\ \bibinfo {author} {\bibfnamefont {C.}~\bibnamefont
  {Krishnan}},\ }in\ \href@noop {} {\emph {\bibinfo {booktitle} {Aqueous
  Solutions of Simple Electrolytes}}}\ (\bibinfo  {publisher} {Springer},\
  \bibinfo {year} {1973})\ pp.\ \bibinfo {pages} {1--118}\BibitemShut {NoStop}%
\bibitem [{\citenamefont {Beck}(2013)}]{Beck:2013gp}%
  \BibitemOpen
  \bibfield  {author} {\bibinfo {author} {\bibfnamefont {T.~L.}\ \bibnamefont
  {Beck}},\ }\href@noop {} {\bibfield  {journal} {\bibinfo  {journal} {Chem.
  Phys. Letts.}\ } (\bibinfo {year} {2013})}\BibitemShut {NoStop}%
\bibitem [{\citenamefont {Chaudhari}\ \emph {et~al.}(2018)\citenamefont
  {Chaudhari}, \citenamefont {Vanagas}, \citenamefont {Pratt},\ and\
  \citenamefont {Rempe}}]{ACR}%
  \BibitemOpen
  \bibfield  {author} {\bibinfo {author} {\bibfnamefont {M.~I.}\ \bibnamefont
  {Chaudhari}}, \bibinfo {author} {\bibfnamefont {J.~M.}\ \bibnamefont
  {Vanagas}}, \bibinfo {author} {\bibfnamefont {L.~R.}\ \bibnamefont {Pratt}},
  \ and\ \bibinfo {author} {\bibfnamefont {S.~B.}\ \bibnamefont {Rempe}},\
  }\href@noop {} {\bibfield  {journal} {\bibinfo  {journal} {Acc. Chem. Res.}\
  } (\bibinfo {year} {2018})}\BibitemShut {NoStop}%
\bibitem [{\citenamefont {Rempe}\ and\ \citenamefont
  {Pratt}(2001)}]{Rempe:2001}%
  \BibitemOpen
  \bibfield  {author} {\bibinfo {author} {\bibfnamefont {S.~B.}\ \bibnamefont
  {Rempe}}\ and\ \bibinfo {author} {\bibfnamefont {L.~R.}\ \bibnamefont
  {Pratt}},\ }\bibfield  {booktitle} {\emph {\bibinfo {booktitle} {Fl. Ph.
  Equ.}},\ }\href@noop {} {\ \textbf {\bibinfo {volume} {183--184}},\ \bibinfo
  {pages} {121} (\bibinfo {year} {2001})}\BibitemShut {NoStop}%
\bibitem [{\citenamefont {Rempe}, \citenamefont {Asthagiri},\ and\
  \citenamefont {Pratt}(2004)}]{Rempe:K}%
  \BibitemOpen
  \bibfield  {author} {\bibinfo {author} {\bibfnamefont {S.~B.}\ \bibnamefont
  {Rempe}}, \bibinfo {author} {\bibfnamefont {D.}~\bibnamefont {Asthagiri}}, \
  and\ \bibinfo {author} {\bibfnamefont {L.~R.}\ \bibnamefont {Pratt}},\
  }\href@noop {} {\bibfield  {journal} {\bibinfo  {journal} {Phys. Chem. Chem.
  Phys.}\ }\textbf {\bibinfo {volume} {6}},\ \bibinfo {pages} {1966} (\bibinfo
  {year} {2004})}\BibitemShut {NoStop}%
\bibitem [{\citenamefont {Asthagiri}\ \emph {et~al.}(2004)\citenamefont
  {Asthagiri}, \citenamefont {Pratt}, \citenamefont {Paulaitis},\ and\
  \citenamefont {Rempe}}]{Asthagiri:divalents}%
  \BibitemOpen
  \bibfield  {author} {\bibinfo {author} {\bibfnamefont {D.}~\bibnamefont
  {Asthagiri}}, \bibinfo {author} {\bibfnamefont {L.~R.}\ \bibnamefont
  {Pratt}}, \bibinfo {author} {\bibfnamefont {M.~E.}\ \bibnamefont
  {Paulaitis}}, \ and\ \bibinfo {author} {\bibfnamefont {S.~B.}\ \bibnamefont
  {Rempe}},\ }\href@noop {} {\bibfield  {journal} {\bibinfo  {journal} {J. Am.
  Chem. Soc.}\ }\textbf {\bibinfo {volume} {126}},\ \bibinfo {pages} {1285}
  (\bibinfo {year} {2004})}\BibitemShut {NoStop}%
\bibitem [{\citenamefont {Jiao}\ \emph {et~al.}(2011)\citenamefont {Jiao},
  \citenamefont {Leung}, \citenamefont {Rempe},\ and\ \citenamefont
  {Nenoff}}]{Jiao:2011}%
  \BibitemOpen
  \bibfield  {author} {\bibinfo {author} {\bibfnamefont {D.}~\bibnamefont
  {Jiao}}, \bibinfo {author} {\bibfnamefont {K.}~\bibnamefont {Leung}},
  \bibinfo {author} {\bibfnamefont {S.~B.}\ \bibnamefont {Rempe}}, \ and\
  \bibinfo {author} {\bibfnamefont {T.~M.}\ \bibnamefont {Nenoff}},\
  }\href@noop {} {\bibfield  {journal} {\bibinfo  {journal} {J. Chem. Theo.
  Comp.}\ }\textbf {\bibinfo {volume} {7}},\ \bibinfo {pages} {485} (\bibinfo
  {year} {2011})}\BibitemShut {NoStop}%
\bibitem [{\citenamefont {Alam}, \citenamefont {Hart},\ and\ \citenamefont
  {Rempe}(2011)}]{Alam:Li}%
  \BibitemOpen
  \bibfield  {author} {\bibinfo {author} {\bibfnamefont {T.~M.}\ \bibnamefont
  {Alam}}, \bibinfo {author} {\bibfnamefont {D.}~\bibnamefont {Hart}}, \ and\
  \bibinfo {author} {\bibfnamefont {S.~L.~B.}\ \bibnamefont {Rempe}},\
  }\href@noop {} {\bibfield  {journal} {\bibinfo  {journal} {Phys. Chem. Chem.
  Phys.}\ }\textbf {\bibinfo {volume} {13}},\ \bibinfo {pages} {13629}
  (\bibinfo {year} {2011})}\BibitemShut {NoStop}%
\bibitem [{\citenamefont {Sabo}\ \emph {et~al.}(2013)\citenamefont {Sabo},
  \citenamefont {Jiao}, \citenamefont {Varma}, \citenamefont {Pratt},\ and\
  \citenamefont {Rempe}}]{Sabo:2013gs}%
  \BibitemOpen
  \bibfield  {author} {\bibinfo {author} {\bibfnamefont {D.}~\bibnamefont
  {Sabo}}, \bibinfo {author} {\bibfnamefont {D.}~\bibnamefont {Jiao}}, \bibinfo
  {author} {\bibfnamefont {S.}~\bibnamefont {Varma}}, \bibinfo {author}
  {\bibfnamefont {L.~R.}\ \bibnamefont {Pratt}}, \ and\ \bibinfo {author}
  {\bibfnamefont {S.~B.}\ \bibnamefont {Rempe}},\ }\href@noop {} {\bibfield
  {journal} {\bibinfo  {journal} {Ann. Rep. Prog. Chem, Sect. C (Phys. Chem.)}\
  }\textbf {\bibinfo {volume} {109}},\ \bibinfo {pages} {266} (\bibinfo {year}
  {2013})}\BibitemShut {NoStop}%
\bibitem [{\citenamefont {Mason}\ \emph {et~al.}(2015)\citenamefont {Mason},
  \citenamefont {Ansell}, \citenamefont {Neilson},\ and\ \citenamefont
  {Rempe}}]{Mason:Li}%
  \BibitemOpen
  \bibfield  {author} {\bibinfo {author} {\bibfnamefont {P.~E.}\ \bibnamefont
  {Mason}}, \bibinfo {author} {\bibfnamefont {S.}~\bibnamefont {Ansell}},
  \bibinfo {author} {\bibfnamefont {G.~W.}\ \bibnamefont {Neilson}}, \ and\
  \bibinfo {author} {\bibfnamefont {S.~B.}\ \bibnamefont {Rempe}},\ }\href@noop
  {} {\bibfield  {journal} {\bibinfo  {journal} {J. Phys. Chem. B}\ }\textbf
  {\bibinfo {volume} {119}},\ \bibinfo {pages} {2003} (\bibinfo {year}
  {2015})}\BibitemShut {NoStop}%
\bibitem [{\citenamefont {Chaudhari}, \citenamefont {Soniat},\ and\
  \citenamefont {Rempe}(2015)}]{Chaudhari:Ba}%
  \BibitemOpen
  \bibfield  {author} {\bibinfo {author} {\bibfnamefont {M.~I.}\ \bibnamefont
  {Chaudhari}}, \bibinfo {author} {\bibfnamefont {M.}~\bibnamefont {Soniat}}, \
  and\ \bibinfo {author} {\bibfnamefont {S.~B.}\ \bibnamefont {Rempe}},\
  }\href@noop {} {\bibfield  {journal} {\bibinfo  {journal} {J. Phys. Chem. B}\
  }\textbf {\bibinfo {volume} {119}},\ \bibinfo {pages} {8746} (\bibinfo {year}
  {2015})}\BibitemShut {NoStop}%
\bibitem [{\citenamefont {Varma}\ and\ \citenamefont
  {Rempe}(2007)}]{Varma:2007ej}%
  \BibitemOpen
  \bibfield  {author} {\bibinfo {author} {\bibfnamefont {S.}~\bibnamefont
  {Varma}}\ and\ \bibinfo {author} {\bibfnamefont {S.~B.}\ \bibnamefont
  {Rempe}},\ }\href@noop {} {\bibfield  {journal} {\bibinfo  {journal}
  {Biophys. J.}\ }\textbf {\bibinfo {volume} {93}},\ \bibinfo {pages} {1093}
  (\bibinfo {year} {2007})}\BibitemShut {NoStop}%
\bibitem [{\citenamefont {Varma}, \citenamefont {Sabo},\ and\ \citenamefont
  {Rempe}(2008)}]{Varma:2008kl}%
  \BibitemOpen
  \bibfield  {author} {\bibinfo {author} {\bibfnamefont {S.}~\bibnamefont
  {Varma}}, \bibinfo {author} {\bibfnamefont {D.}~\bibnamefont {Sabo}}, \ and\
  \bibinfo {author} {\bibfnamefont {S.~B.}\ \bibnamefont {Rempe}},\ }\href@noop
  {} {\bibfield  {journal} {\bibinfo  {journal} {J. Mol. Bio.}\ }\textbf
  {\bibinfo {volume} {376}},\ \bibinfo {pages} {13} (\bibinfo {year}
  {2008})}\BibitemShut {NoStop}%
\bibitem [{\citenamefont {Varma}\ \emph {et~al.}(2011)\citenamefont {Varma},
  \citenamefont {Rogers}, \citenamefont {Pratt},\ and\ \citenamefont
  {Rempe}}]{Varma:2011ho}%
  \BibitemOpen
  \bibfield  {author} {\bibinfo {author} {\bibfnamefont {S.}~\bibnamefont
  {Varma}}, \bibinfo {author} {\bibfnamefont {D.~M.}\ \bibnamefont {Rogers}},
  \bibinfo {author} {\bibfnamefont {L.~R.}\ \bibnamefont {Pratt}}, \ and\
  \bibinfo {author} {\bibfnamefont {S.~B.}\ \bibnamefont {Rempe}},\ }\href
  {\doibase 10.1085/jgp.201010579} {\bibfield  {journal} {\bibinfo  {journal}
  {J. Gen. Phys.}\ }\textbf {\bibinfo {volume} {137}},\ \bibinfo {pages} {479}
  (\bibinfo {year} {2011})}\BibitemShut {NoStop}%
\bibitem [{\citenamefont {Rossi}\ \emph {et~al.}(2013)\citenamefont {Rossi},
  \citenamefont {Tkatchenko}, \citenamefont {Rempe},\ and\ \citenamefont
  {Varma}}]{Rossi:2013gm}%
  \BibitemOpen
  \bibfield  {author} {\bibinfo {author} {\bibfnamefont {M.}~\bibnamefont
  {Rossi}}, \bibinfo {author} {\bibfnamefont {A.}~\bibnamefont {Tkatchenko}},
  \bibinfo {author} {\bibfnamefont {S.~B.}\ \bibnamefont {Rempe}}, \ and\
  \bibinfo {author} {\bibfnamefont {S.}~\bibnamefont {Varma}},\ }\href@noop {}
  {\bibfield  {journal} {\bibinfo  {journal} {Proc. Natl. Acad. Sci. U.S.A.}\
  }\textbf {\bibinfo {volume} {110}},\ \bibinfo {pages} {12978} (\bibinfo
  {year} {2013})}\BibitemShut {NoStop}%
\bibitem [{\citenamefont {Stevens}\ and\ \citenamefont
  {Rempe}(2016)}]{Stevens}%
  \BibitemOpen
  \bibfield  {author} {\bibinfo {author} {\bibfnamefont {M.}~\bibnamefont
  {Stevens}}\ and\ \bibinfo {author} {\bibfnamefont {S.~B.}\ \bibnamefont
  {Rempe}},\ }\href@noop {} {\bibfield  {journal} {\bibinfo  {journal} {J.
  Phys. Chem. B}\ }\textbf {\bibinfo {volume} {120}},\ \bibinfo {pages} {12519}
  (\bibinfo {year} {2016})}\BibitemShut {NoStop}%
\bibitem [{\citenamefont {Chaudhari}\ and\ \citenamefont
  {Rempe}(2018)}]{Chaudhari:Sr}%
  \BibitemOpen
  \bibfield  {author} {\bibinfo {author} {\bibfnamefont {M.~I.}\ \bibnamefont
  {Chaudhari}}\ and\ \bibinfo {author} {\bibfnamefont {S.~B.}\ \bibnamefont
  {Rempe}},\ }\href@noop {} {\bibfield  {journal} {\bibinfo  {journal} {J.
  Chem. Phys.}\ ,\ \bibinfo {pages} {222831}} (\bibinfo {year}
  {2018})}\BibitemShut {NoStop}%
\bibitem [{\citenamefont {Chaudhari}, \citenamefont {Rempe},\ and\
  \citenamefont {Pratt}(2017)}]{chaudhari2017quasi}%
  \BibitemOpen
  \bibfield  {author} {\bibinfo {author} {\bibfnamefont {M.~I.}\ \bibnamefont
  {Chaudhari}}, \bibinfo {author} {\bibfnamefont {S.~B.}\ \bibnamefont
  {Rempe}}, \ and\ \bibinfo {author} {\bibfnamefont {L.~R.}\ \bibnamefont
  {Pratt}},\ }\href@noop {} {\bibfield  {journal} {\bibinfo  {journal} {J.
  Chem. Phys.}\ }\textbf {\bibinfo {volume} {147}},\ \bibinfo {pages} {161728}
  (\bibinfo {year} {2017})}\BibitemShut {NoStop}%
\bibitem [{\citenamefont {Tissandier}\ \emph {et~al.}(1998)\citenamefont
  {Tissandier}, \citenamefont {Cowen}, \citenamefont {Feng}, \citenamefont
  {Gundlach}, \citenamefont {Cohen}, \citenamefont {Earhart}, \citenamefont
  {Coe},\ and\ \citenamefont {Tuttle}}]{tissandier1998proton}%
  \BibitemOpen
  \bibfield  {author} {\bibinfo {author} {\bibfnamefont {M.~D.}\ \bibnamefont
  {Tissandier}}, \bibinfo {author} {\bibfnamefont {K.~A.}\ \bibnamefont
  {Cowen}}, \bibinfo {author} {\bibfnamefont {W.~Y.}\ \bibnamefont {Feng}},
  \bibinfo {author} {\bibfnamefont {E.}~\bibnamefont {Gundlach}}, \bibinfo
  {author} {\bibfnamefont {M.~H.}\ \bibnamefont {Cohen}}, \bibinfo {author}
  {\bibfnamefont {A.~D.}\ \bibnamefont {Earhart}}, \bibinfo {author}
  {\bibfnamefont {J.~V.}\ \bibnamefont {Coe}}, \ and\ \bibinfo {author}
  {\bibfnamefont {T.~R.}\ \bibnamefont {Tuttle}},\ }\href@noop {} {\bibfield
  {journal} {\bibinfo  {journal} {J. Phys. Chem. A}\ }\textbf {\bibinfo
  {volume} {102}},\ \bibinfo {pages} {7787} (\bibinfo {year}
  {1998})}\BibitemShut {NoStop}%
\bibitem [{\citenamefont {Tomasi}, \citenamefont {Mennucci},\ and\
  \citenamefont {Cammi}(2005)}]{Tomasi:2005tc}%
  \BibitemOpen
  \bibfield  {author} {\bibinfo {author} {\bibfnamefont {J.}~\bibnamefont
  {Tomasi}}, \bibinfo {author} {\bibfnamefont {B.}~\bibnamefont {Mennucci}}, \
  and\ \bibinfo {author} {\bibfnamefont {R.}~\bibnamefont {Cammi}},\
  }\href@noop {} {\bibfield  {journal} {\bibinfo  {journal} {Chem. Rev.}\
  }\textbf {\bibinfo {volume} {105}},\ \bibinfo {pages} {2999} (\bibinfo {year}
  {2005})}\BibitemShut {NoStop}%
\bibitem [{\citenamefont {Frisch}\ \emph {et~al.}(1998)\citenamefont {Frisch},
  \citenamefont {Trucks}, \citenamefont {Schlegel}, \citenamefont {Scuseria},
  \citenamefont {Robb}, \citenamefont {Cheeseman}, \citenamefont {Zakrzewski},
  \citenamefont {Montgomery~Jr}, \citenamefont {Stratmann}, \citenamefont
  {Burant} \emph {et~al.}}]{frisch1998gaussian}%
  \BibitemOpen
  \bibfield  {author} {\bibinfo {author} {\bibfnamefont {M.}~\bibnamefont
  {Frisch}}, \bibinfo {author} {\bibfnamefont {G.}~\bibnamefont {Trucks}},
  \bibinfo {author} {\bibfnamefont {H.}~\bibnamefont {Schlegel}}, \bibinfo
  {author} {\bibfnamefont {G.}~\bibnamefont {Scuseria}}, \bibinfo {author}
  {\bibfnamefont {M.}~\bibnamefont {Robb}}, \bibinfo {author} {\bibfnamefont
  {J.}~\bibnamefont {Cheeseman}}, \bibinfo {author} {\bibfnamefont
  {V.}~\bibnamefont {Zakrzewski}}, \bibinfo {author} {\bibfnamefont
  {J.}~\bibnamefont {Montgomery~Jr}}, \bibinfo {author} {\bibfnamefont
  {R.}~\bibnamefont {Stratmann}}, \bibinfo {author} {\bibfnamefont
  {J.}~\bibnamefont {Burant}},  \emph {et~al.},\ }\href@noop {} {\bibfield
  {journal} {\bibinfo  {journal} {Pittsburgh, Pa}\ }\textbf {\bibinfo {volume}
  {40}} (\bibinfo {year} {1998})}\BibitemShut {NoStop}%
\bibitem [{\citenamefont {Perdew}, \citenamefont {Burke},\ and\ \citenamefont
  {Ernzerhof}(1996)}]{perdew1996generalized}%
  \BibitemOpen
  \bibfield  {author} {\bibinfo {author} {\bibfnamefont {J.~P.}\ \bibnamefont
  {Perdew}}, \bibinfo {author} {\bibfnamefont {K.}~\bibnamefont {Burke}}, \
  and\ \bibinfo {author} {\bibfnamefont {M.}~\bibnamefont {Ernzerhof}},\
  }\href@noop {} {\bibfield  {journal} {\bibinfo  {journal} {Phys. Rev.
  Letts.}\ }\textbf {\bibinfo {volume} {77}},\ \bibinfo {pages} {3865}
  (\bibinfo {year} {1996})}\BibitemShut {NoStop}%
\bibitem [{\citenamefont {Becke}(1993)}]{becke1993density}%
  \BibitemOpen
  \bibfield  {author} {\bibinfo {author} {\bibfnamefont {A.~D.}\ \bibnamefont
  {Becke}},\ }\href@noop {} {\bibfield  {journal} {\bibinfo  {journal} {J.
  Chem. Phys.}\ }\textbf {\bibinfo {volume} {98}},\ \bibinfo {pages} {5648}
  (\bibinfo {year} {1993})}\BibitemShut {NoStop}%
\bibitem [{\citenamefont {Lee}, \citenamefont {Yang},\ and\ \citenamefont
  {Parr}(1988)}]{lee1988development}%
  \BibitemOpen
  \bibfield  {author} {\bibinfo {author} {\bibfnamefont {C.}~\bibnamefont
  {Lee}}, \bibinfo {author} {\bibfnamefont {W.}~\bibnamefont {Yang}}, \ and\
  \bibinfo {author} {\bibfnamefont {R.~G.}\ \bibnamefont {Parr}},\ }\href@noop
  {} {\bibfield  {journal} {\bibinfo  {journal} {Phys. Rev. B}\ }\textbf
  {\bibinfo {volume} {37}},\ \bibinfo {pages} {785} (\bibinfo {year}
  {1988})}\BibitemShut {NoStop}%
\bibitem [{\citenamefont {Dunning~Jr}(1989)}]{dunning1989gaussian}%
  \BibitemOpen
  \bibfield  {author} {\bibinfo {author} {\bibfnamefont {T.~H.}\ \bibnamefont
  {Dunning~Jr}},\ }\href@noop {} {\bibfield  {journal} {\bibinfo  {journal} {J.
  Chem. Phys.}\ }\textbf {\bibinfo {volume} {90}},\ \bibinfo {pages} {1007}
  (\bibinfo {year} {1989})}\BibitemShut {NoStop}%
\bibitem [{\citenamefont {Rempe}\ and\ \citenamefont
  {J{\'o}nsson}(1998)}]{Rempe:normal}%
  \BibitemOpen
  \bibfield  {author} {\bibinfo {author} {\bibfnamefont {S.~B.}\ \bibnamefont
  {Rempe}}\ and\ \bibinfo {author} {\bibfnamefont {H.}~\bibnamefont
  {J{\'o}nsson}},\ }\href@noop {} {\bibfield  {journal} {\bibinfo  {journal}
  {Chem. Educator}\ }\textbf {\bibinfo {volume} {3}},\ \bibinfo {pages} {1}
  (\bibinfo {year} {1998})}\BibitemShut {NoStop}%
\bibitem [{\citenamefont {Ochterski}(2000)}]{Ochterski2000thermochemistry}%
  \BibitemOpen
  \bibfield  {author} {\bibinfo {author} {\bibfnamefont {J.~W.}\ \bibnamefont
  {Ochterski}},\ }\href@noop {} {\emph {\bibinfo {title} {Thermochemistry in
  gaussian}}}\ (\bibinfo  {publisher} {Gaussian Inc},\ \bibinfo {year}
  {2000})\BibitemShut {NoStop}%
\bibitem [{\citenamefont {Irikura}(1998)}]{Irikura:1998te}%
  \BibitemOpen
  \bibfield  {author} {\bibinfo {author} {\bibfnamefont {K.~K.}\ \bibnamefont
  {Irikura}},\ }\href@noop {} {\emph {\bibinfo {title} {{Computational
  Thermochemistry Prediction and Estimation of Molecular Thermodynamics}}}},\
  edited by\ \bibinfo {editor} {\bibfnamefont {K.~K.}\ \bibnamefont {Irikura}}\
  and\ \bibinfo {editor} {\bibfnamefont {D.~J.}\ \bibnamefont {Frurip}},\ Vol.\
  \bibinfo {volume} {677}\ (\bibinfo  {publisher} {ACS Symposium Series},\
  \bibinfo {year} {1998})\BibitemShut {NoStop}%
\bibitem [{\citenamefont {Kresse}\ and\ \citenamefont
  {Hafner}(1993)}]{kresse1993ab}%
  \BibitemOpen
  \bibfield  {author} {\bibinfo {author} {\bibfnamefont {G.}~\bibnamefont
  {Kresse}}\ and\ \bibinfo {author} {\bibfnamefont {J.}~\bibnamefont
  {Hafner}},\ }\href@noop {} {\bibfield  {journal} {\bibinfo  {journal}
  {Physical Review B}\ }\textbf {\bibinfo {volume} {47}},\ \bibinfo {pages}
  {558} (\bibinfo {year} {1993})}\BibitemShut {NoStop}%
\bibitem [{\citenamefont {Kresse}\ and\ \citenamefont
  {Furthm{\"u}ller}(1996)}]{kresse1996efficient}%
  \BibitemOpen
  \bibfield  {author} {\bibinfo {author} {\bibfnamefont {G.}~\bibnamefont
  {Kresse}}\ and\ \bibinfo {author} {\bibfnamefont {J.}~\bibnamefont
  {Furthm{\"u}ller}},\ }\href@noop {} {\bibfield  {journal} {\bibinfo
  {journal} {Physical review B}\ }\textbf {\bibinfo {volume} {54}},\ \bibinfo
  {pages} {11169} (\bibinfo {year} {1996})}\BibitemShut {NoStop}%
\bibitem [{\citenamefont {Dullens}\ \emph {et~al.}(2005)\citenamefont
  {Dullens}, \citenamefont {Aarts}, \citenamefont {Kegel},\ and\ \citenamefont
  {Lekkerkerker}}]{dullens2005widom}%
  \BibitemOpen
  \bibfield  {author} {\bibinfo {author} {\bibfnamefont {R.~P.}\ \bibnamefont
  {Dullens}}, \bibinfo {author} {\bibfnamefont {D.~G.}\ \bibnamefont {Aarts}},
  \bibinfo {author} {\bibfnamefont {W.~K.}\ \bibnamefont {Kegel}}, \ and\
  \bibinfo {author} {\bibfnamefont {H.~N.}\ \bibnamefont {Lekkerkerker}},\
  }\href@noop {} {\bibfield  {journal} {\bibinfo  {journal} {Mol. Phys.}\
  }\textbf {\bibinfo {volume} {103}},\ \bibinfo {pages} {3195} (\bibinfo {year}
  {2005})}\BibitemShut {NoStop}%
\bibitem [{\citenamefont {Soniat}, \citenamefont {Rogers},\ and\ \citenamefont
  {Rempe}(2015)}]{Soniat:2015kl}%
  \BibitemOpen
  \bibfield  {author} {\bibinfo {author} {\bibfnamefont {M.}~\bibnamefont
  {Soniat}}, \bibinfo {author} {\bibfnamefont {D.~M.}\ \bibnamefont {Rogers}},
  \ and\ \bibinfo {author} {\bibfnamefont {S.~B.}\ \bibnamefont {Rempe}},\
  }\href@noop {} {\bibfield  {journal} {\bibinfo  {journal} {J. Chem. Theory
  Comput.}\ }\textbf {\bibinfo {volume} {11}},\ \bibinfo {pages}
  {150626121718001} (\bibinfo {year} {2015})}\BibitemShut {NoStop}%
\bibitem [{\citenamefont {Kathmann}, \citenamefont {Schenter},\ and\
  \citenamefont {Garrett}(2007)}]{kathmann2007critical}%
  \BibitemOpen
  \bibfield  {author} {\bibinfo {author} {\bibfnamefont {S.}~\bibnamefont
  {Kathmann}}, \bibinfo {author} {\bibfnamefont {G.}~\bibnamefont {Schenter}},
  \ and\ \bibinfo {author} {\bibfnamefont {B.}~\bibnamefont {Garrett}},\
  }\href@noop {} {\bibfield  {journal} {\bibinfo  {journal} {J. Phys. Chem. C}\
  }\textbf {\bibinfo {volume} {111}},\ \bibinfo {pages} {4977} (\bibinfo {year}
  {2007})}\BibitemShut {NoStop}%
\bibitem [{\citenamefont {Marcus}(1994)}]{Marcus:1994ci}%
  \BibitemOpen
  \bibfield  {author} {\bibinfo {author} {\bibfnamefont {Y.}~\bibnamefont
  {Marcus}},\ }\href {\doibase 10.1016/0301-4622(94)00051-4} {\bibfield
  {journal} {\bibinfo  {journal} {Biophys. Chem.}\ }\textbf {\bibinfo {volume}
  {51}},\ \bibinfo {pages} {111} (\bibinfo {year} {1994})}\BibitemShut
  {NoStop}%
\bibitem [{\citenamefont {H{\"u}nenberger}\ and\ \citenamefont
  {Reif}(2011)}]{hunenberger2011single}%
  \BibitemOpen
  \bibfield  {author} {\bibinfo {author} {\bibfnamefont {P.}~\bibnamefont
  {H{\"u}nenberger}}\ and\ \bibinfo {author} {\bibfnamefont {M.}~\bibnamefont
  {Reif}},\ }\href@noop {} {\emph {\bibinfo {title} {Single-Ion Solvation:
  Experimental and Theoretical Approaches to Elusive Thermodynamic
  Quantities}}},\ Vol.~\bibinfo {volume} {3}\ (\bibinfo  {publisher} {Royal
  Society of Chemistry},\ \bibinfo {year} {2011})\BibitemShut {NoStop}%
\bibitem [{\citenamefont {You}, \citenamefont {Chaudhari},\ and\ \citenamefont
  {Pratt}(2014)}]{you2014comparison}%
  \BibitemOpen
  \bibfield  {author} {\bibinfo {author} {\bibfnamefont {X.}~\bibnamefont
  {You}}, \bibinfo {author} {\bibfnamefont {M.}~\bibnamefont {Chaudhari}}, \
  and\ \bibinfo {author} {\bibfnamefont {L.}~\bibnamefont {Pratt}},\ }in\
  \href@noop {} {\emph {\bibinfo {booktitle} {Aqua Incognita: Why Ice Floats on
  Water and Galileo 400 Years on}}}\ (\bibinfo  {publisher} {Connor Court Press
  Ballarat},\ \bibinfo {year} {2014})\ pp.\ \bibinfo {pages} {434--442},\
  \bibinfo {note} {{C}omparison of Mechanical and Thermodynamical Evaluations
  of Electrostatic Potential Differences between Electrolyte
  Solutions}\BibitemShut {NoStop}%
\end{thebibliography}%
%merlin.mbs aipnum4-1.bst 2010-07-25 4.21a (PWD, AO, DPC) hacked
%Control: key (0)
%Control: author (8) initials jnrlst
%Control: editor formatted (1) identically to author
%Control: production of article title (-1) disabled
%Control: page (0) single
%Control: year (1) truncated
%Control: production of eprint (0) enabled
%

\end{document}